\documentclass[12pt]{article}
\usepackage{amsfonts, amsmath, amssymb, cite, graphicx, }

\usepackage{color}
\usepackage[all, knot]{xy}
\usepackage{tikz}
\usepackage{subfigure}
\usepackage{braket}
\usepackage{appendix}
\usepackage{makecell}

\usepackage[utf8]{inputenc}

\usepackage{epstopdf}
\usepackage[footnotesize]{caption}
\usepackage{amsthm}
\usepackage{enumitem}
\usepackage{mathrsfs}

\usepackage[margin=3cm]{geometry}

\def \be {\begin{equation}}
\def \ee {\end{equation}}
\def \ba {\begin{aligned}}
\def \ea {\end{aligned}}
\def \bea {\begin{eqnarray}}
\def \eea {\end{eqnarray}}

\begin{document}

\begin{titlepage}
\vspace{0.5cm}
\begin{center}
{\Large \bf The topological natures of the Gauss-Bonnet black hole in AdS space}

\lineskip .75em
\vskip 1cm
{\large Conghua Liu$^{a,b}$, Jin Wang$^{c,}$\footnote{Corresponding author.\\ jin.wang.1@stonybrook.edu} }
\vskip 2.5em
 {\normalsize\it $^{a}$College of Physics, Jilin University, Changchun 130022, China\\
 $^{b}$State Key Laboratory of Electroanalytical Chemistry, Changchun Institute of Applied Chemistry, Chinese Academy of Sciences, Changchun 130022, China\\
 $^{c}$Department of Chemistry and Department of Physics and Astronomy, State University of New York at Stony Brook, Stony Brook, New York 11794, U.S.A}
\vskip 3.0em
\end{center}
\begin{abstract}
In the recent proposal [Phys. Rev. Lett. {\bf 129}, 191101 (2022)], the black holes were viewed as topological thermodynamic defects by using the generalized off-shell free energy. In this paper, we follow such proposal to study the local and global topological natures of the Gauss-Bonnet black holes in AdS space. The local topological natures are reflected by the winding numbers, where the positive and negative winding numbers correspond to the stable and unstable black hole branches. The global topological natures are reflected by the topological numbers, which are defined as the sum of the winding numbers for all black hole branches and can be used to classify the black holes into different classes. When the charge is present, we find that the topological number is independent on the values of the parameters, and the charged Gauss-Bonnet AdS black holes can be divided into the same class of the RNAdS black holes with the same topological number $1$. However, when the charge is absent, we find that the topological number has certain dimensional dependence. This is different from the previous studies, where the topological number is found to be a universal number independent of the black hole parameters. Furthermore, the asymptotic behaviours of curve $\tau(r_h)$ in small and large radii limit can be a simple criterion to distinguish the different topological number. We find a new asymptotic behaviour as $\tau(r_h\to 0)=0$ and $\tau(r_h\to\infty)=0$ in the black hole system, which shows topological equivalency with the asymptotic behaviours $\tau(r_h\to 0)=\infty$ and $\tau(r_h\to\infty)=\infty$. We also give an intuitional proof of why there are only three topological classes in the black hole system under the condition $(\partial_{r_h}S)_P>0$.

 \end{abstract}
\end{titlepage}

\baselineskip=0.7cm

\tableofcontents
\newpage
\section{Introduction}

Since the pioneering works proposed in~\cite{AA,AB,AC,AD}, the black hole thermodynamics has made significant progress in recent years. A famous example is the Hawking-page phase transition occurring in the AdS space, where a first order phase transition has been found between the large SAdS black hole and the thermal radiation~\cite{AE}. The subsequent researches showed that the Hawking-Page phase transition can be interpreted as the confinement/deconfinement phase transition in the context of AdS/CFT correspondence~\cite{AF}. In the canonical ensemble with the fixed charge, the free energy of charged AdS black hole shows a swallowtail behavior, which means a first order phase transition can also occur between the small black hole (SBH) and the large black hole (LBH)~\cite{AG,AH}. Furthermore, the $Q-\Phi$ (charge-chemical potential) diagram of the charged AdS black holes was found to be similar to the $P-v$ (pressure-volume) diagram of the vdW fluids~\cite{AG,AH}.

Recently, the remarkable analogies between the charged AdS black hole and the van de Waals fluids have been restudied in the extended phase space~\cite{AI,AJ,AK,AL,AM,AN}. In the extended phase space, the cosmological constant plays the role of the thermodynamic pressure whose conjugate quantity is the thermodynamic volume, and the black hole mass is interpreted as the enthalpy rather than the internal energy~\cite{AI,AJ,AK}. After treating the cosmological constant as the thermodynamic pressure, the first law of black hole thermodynamics should involve the cosmological constant, and the consistent Smarr relation was held~\cite{AI,AJ,AK}. It was found that the charged AdS black hole and the vdW fluids not only behave analogously in the phase diagram, but also have the similar equations of state and the same critical exponents~\cite{AL,AM}. Furthermore, the mesoscopic structure revealed by the radial distribution function were found to be same between the small/large charged AdS black holes and the liquid/gas systems~\cite{AO}.

The phase behaviours have also been studied in the extended phase space of Gauss-Bonnet black holes in the AdS space when $d \geqslant 5$ dimensions~\cite{AP,AQ}. When the geometry of the black hole horizon is Ricci flat ($k=0$) or hyperbolic ($k=-1$), it was found that there was no critical point and thus no phase transition to occur. Only when the geometry of the black hole horizon is spherical ($k=1$), the phase transition can take place. However, in the case of spherical horizon,  the charged Gauss-Bonnet black holes show different phase behaviours in different dimensions. When $d=5$ or $d \geqslant 7$, a small/large black hole (SBH/LBH) phase transition which is similar to that of the vdW fluids has been found. When $d=6$, the phase behaviours of the charged Gauss-Bonnet black holes are extremely abundant. There is not only the SBH/LBH phase transition which behaves like that of the vdW fluids, but also the SBH/IBH/LBH phase transition which behaves like the solid/liquid/gas phase transition, and the triple point can be found~\cite{AQ}.

The free energy $G$ can be generalized from the standard definition of the free energy $G=H-T_HS$, where the enthalpy $H$ represents the mass of the black hole in the extended phase space, $S$ is the entropy and $T_H$ is the Hawking temperature~\cite{AR,AS,AT}. Recently, the concept of generalized free energy has been extended to the five-dimensional Schwarzschild black holes in the canonical ensemble~\cite{AS}. In particular, the generalized free energy can be derived from the Einstein-Hilbert action of the Euclidean gravitational instanton with the conical singularity~\cite{AT}. In the free energy landscape, one can consider a canonical ensemble composed of various black hole states with different radii at fixed temperature. The Hawking temperature $T_H$ is then replaced by the ensemble temperature $T$, and the free energy becomes a continuous function of the black hole radius. Namely, the free energy is defined not only on the on-shell states which satisfy the stationary Einstein field equation, but also on the off-shell states which do not obey the stationary Einstein field equation. However, the extremal points of the free energy landscape exactly correspond to the on-shell black holes. Based on the free energy landscape, one can have an intuitive picture on the thermodynamic stability of different on-shell black holes at fixed temperature. Namely, the local minimum or maximum point on the free energy landscape represents the local stable and unstable black holes, and the global minimum point represents the global stable black hole. Recently, the researches show that the free energy landscape is not only beneficial for the black hole thermodynamics, but also can be used to analyze the dynamics of the black hole phase transition, where the off-shell states play a vital role in revealing the dynamical process during the phase transition~\cite{AU,AV,AW}.

In~\cite{AX,AY}, a special vector was constructed by means of the null geodesics, where the light ring of the black hole is located at the zero point of the vector field. The topological argument shows that there exists at least one standard light ring outside the black hole horizon for each rotation sense of the four-dimensional stationary, axisymmetric, asymptotically flat black hole with a nonextremal, topologically spherical Killing horizon. Following such a topological argument and the topological current $\phi$-mapping theory, each critical point of the phase transition was associated with a topological charge and the critical points were divided into two classes~\cite{AZ,BA,BB,BC}. On the other hand, the black hole solutions were treated as defects in the thermodynamic parameter space by means of the generalized off-shell free energy~\cite{BF}. Namely, a new vector was constructed where the on-shell black holes were located exactly at the zero point of the vector field. Following the similar methods of the topological classification of the critical points, the positive and negative winding numbers corresponding to the defect indicated the local thermodynamical stable and unstable black hole solutions, respectively. Furthermore, the topological number is defined as the sum of the winding numbers for all black hole branches, which can be used to classify the black holes to different classes. Recently, the topology in black hole thermodynamics has attracted a lot of attention and been used in different black holes~\cite{BD,BE,BL,QA,QB,QC,QD,QE,QF,QG}

In this paper, we study the topological classification of the Gauss-Bonnet AdS black with different geometries of the black hole horizons in different dimensions. The motivations are as follows. (1) The geometry of the black hole horizon in AdS space can be Ricci flat, spherical or hyperbolic, how would that change the topological number? (2) When the black hole horizon is spherical, the phase behaviors of the charged Gauss-Bonnet AdS black in $d=6$ dimensions are very different from those in other dimensions, how would that change the topological number? (3) How would the higher derivative terms of curvature in the form of Gauss-Bonnet gravity change the topology number? (4) The topological numbers are different from the Schwarzschild black holes to the RN black holes, how would the charge change the topological number of the Gauss-Bonnet AdS black holes? 
 
The paper is organized as follows. In section~\ref{TAT}, we introduce the thermodynamics of the Gauss-Bonnet AdS black hole and the topology. In section~\ref{WCP}, we study the topological natures of the Gauss-Bonnet AdS black holes when the charge is present, where the cases of Ricci flat, hyperbolic and spherical horizons are discussed respectively. We find the topological number to be independent on the black hole parameter when the charge is present, and the charged Gauss-Bonnet AdS black holes can be divided into the same class of the RNAdS black holes. In section~\ref{WCA}, the topological properties will be restudied when the charge is absent. When the geometry of the horizon is spherical, a new asymptotic behaviour of curves $\tau(r_h)$ was found in $d\geq 6$, whose topological number is different from that in $d=5$. In~\cite{BF}, the topological number was found to be a universal number independent on the black hole parameter. However, our results show the topological number has dimensional dependence in certain case. In section~\ref{WTTC}, we will present an intuitional proof of why there are only three classes of black holes under the condition $(\partial_{r_h}S)_P>0$. In section~\ref{CON}, we present the conclusions.


\section{The thermodynamics of Gauss-Bonnet AdS black hole and the topology}
\label{TAT}
\subsection{The thermodynamics of d-dimensional Gauss-Bonnet AdS black hole}
Beginning with the metric of the $d$-dimensional GB-AdS black hole~\cite{AP,AQ}:
\be
\ba
\label{eq:1}
ds^2=-f(r)dt^2+f^{-1}(r)dr^2+r^2h_{ij}dx^idx^j,
\ea
\ee
where $h_{ij}dx^idx^j$ represents the line element of a ($d-2$)-dimensional maximal symmetric Einstein space with constant curvature $(d-2)(d-3)k$ and volume $\Sigma_k$. $k=0$, $-1$ or $1$ represents the Ricci flat, hyperbolic and spherical horizons, respectively. The metric function $f(r)$ is given by
\be
\ba
\label{eq:2}
f(r)=k+\frac{r^2}{2\alpha}\{1-\sqrt{1+\frac{64\pi\alpha M}{(d-2)\Sigma_k r^{d-1}}-\frac{2\alpha Q^2}{(d-2)(d-3)r^{2d-4}}-\frac{64\pi\alpha P}{(d-1)(d-2)}}\},
\ea
\ee
where $M$ is the black hole mass, $Q$ is the charge, and $\alpha$ is associated with the positive Gauss-Bonnet coefficient $\alpha_{GB}$ by $\alpha=(d-3)(d-4)\alpha_{GB}$. As discussed in the Introduction, the cosmological constant $\Lambda$ is considered as a dynamical variable in the extended phase space, which is interpreted as the pressure $P$ by $P=-\frac{\Lambda}{8\pi}$ due to the realization that the negative cosmological constant induces a positive vacuum pressure in spacetime. Without loss of generality, we set $\alpha$ and $\Sigma_k$ as $1$ for simplification.

The thermodynamic quantities can be expressed in terms of the horizon radius $r_h$ which is determined by the largest real root of the equation $f(r)=0$, we list the thermodynamic quantities as follows.
\be
\ba
\label{eq:3}
M=\frac{(d-2){r_h}^{d-3}}{16\pi}(k+\frac{k^2}{{r_h}^2}+\frac{16\pi P{r_h}^2}{(d-1)(d-2)})+\frac{Q^2}{8\pi(d-3){r_h}^{d-3}},
\ea
\ee

\be
\ba
\label{eq:4}
T_H=\frac{\frac{16\pi P{r_h}^4}{d-2}+(d-3)k{r_h}^2+(d-5)k^2-\frac{2Q^2}{(d-2){r_h}^{2d-8}}}{4\pi r_h({r_h}^2+2k)},
\ea
\ee

\be
\ba
\label{eq:5}
S=\frac{{r_h}^{d-2}}{4}(1+\frac{2(d-2)k}{(d-4){r_h}^2}).
\ea
\ee

In the extended phase space, the black hole mass $M$ is interpreted as the enthalpy rather than the internal energy, so the Gibbs free energy is given by 
\be
\ba
\label{eq:6}
G=M-T_H S.
\ea
\ee

We should note certain constraints as discussed in~\cite{AP}. The first constraint comes from the metric function~(\ref{eq:2}), where a well-defined vacuum solution with $M=Q=0$ requires
\be
\ba
\label{eq:7}
0\leqslant P \leqslant \frac{(d-1)(d-2)}{64\pi\alpha}.
\ea
\ee
Another constraint comes from the non-negative definiteness of the black hole entropy in equation~(\ref{eq:5}), which requires
\be
\ba
\label{eq:8}
{r_h}^2+2k+\frac{4k}{d-4} \geqslant 0.
\ea
\ee

The heat capacity $C_P$ at fixed pressure $P$ can reflect the local thermodynamic stability, i.e. the positive or negative heat capacity corresponds to the stable and unstable system respectively. The heat capacity $C_P$ is defined as
\be
\ba
\label{eq:9}
C_P=T_H(\frac{\partial S}{\partial T_H})_P=T_H(\frac{\partial_{r_h}S}{\partial_{r_h}T_H})_P.
\ea
\ee
By using the equation of entropy~(\ref{eq:5}), we complete the calculation of $(\partial_{r_h}S)_P$, which gives the result as
\be
\ba
\label{eq:10}
(\frac{\partial S}{\partial r_h})_P=\frac{d-2}{4}{r_h}^{d-5}({r_h}^2+2k).
\ea
\ee
The constrain~(\ref{eq:8}) ensures that the equation ~(\ref{eq:10}) can not be negative. By means of the equation~(\ref{eq:9}), the sign of the heat capacity $C_P$ at fixed pressure is found to be the same as the sign of $(\partial_{r_h} T_H)_P$.

\subsection{The topology}
As discussed in the Introduction, in order to generalize the on-shell free energy to off-shell, we consider a canonical ensemble which is composed of various black hole states with different radii at fixed temperature. The generalized free energy can be obtained by replacing the Hawking temperature $T_H$ with the ensemble temperature $T$ in the equation~(\ref{eq:6}). Following the Euclidean path integral approach~\cite{BG}, the ensemble temperature $T$ is equal to $1/\tau$, where the parameter $\tau$ is the Euclidean time period~\cite{BH,BI,BJ}. The generalized Gibbs free energy can then be given by
\be
\ba
\label{eq:11}
G&=M-\frac{S}{\tau} \\
&=\frac{(d-2){r_h}^{d-3}}{16\pi}(k+\frac{k^2}{{r_h}^2}+\frac{16\pi P{r_h}^2}{(d-1)(d-2)})+\frac{Q^2}{8\pi(d-3){r_h}^{d-3}}-\frac{{r_h}^{d-2}}{4\tau}(1+\frac{2(d-2)k}{(d-4){r_h}^2}).
\ea
\ee
Only when $\tau=\tau_H=1/T_H$, the generalized free energy is on-shell. If $\tau\neq\tau_H$, the subspace of the Euclidean manifold has the geometry of a cone with a non-zero deficit angle $2\pi(1-\tau/\tau_H)$, such a manifold is not regular and has a conical singularity near the horizon. When $\tau=\tau_H$, the deficit angle vanishes, so the geometry becomes a disk and the manifold recovers to be regular~\cite{BH,BI,BJ}. 

In~\cite{BF}, a vector $\phi$ is constructed as
\be
\ba
\label{eq:12}
\phi=(\frac{\partial G}{\partial r_h},-\cot\Theta \csc\Theta),
\ea
\ee
where $0\leqslant \Theta \leqslant \pi$ and $0 \leqslant r_h \leqslant \infty$. The introduction of the new parameter $\Theta$ is for the purpose of axial limit, where the component $\phi^{\Theta}$ is divergent and the direction of the vector points outward when $\Theta=0$ and $\pi$~\cite{BD,BE,BF}. Then, the topological number $W$ is determined by the direction of the vector $\phi$ at $r_h=0$ and $\infty$~\cite{BD,BE,BF}. Furthermore, as discussed in the Introduction, the extremal points of the free energy landscape exactly correspond to the on-shell black holes, so the zero point of the component $\phi^{r_h}$ exactly meets the black hole solution. The component $\phi^{\Theta}=0$ will yield $\Theta=\pi/2$.

By using the vector $\phi$, a topological current can be introduced as~\cite{AZ,BA,BB,BF}
\be
\ba
\label{eq:13}
j^{\mu}=\frac{1}{2\pi}{\epsilon}^{\mu\nu\rho}{\epsilon}_{ab}\partial_{\nu}n^a\partial_{\rho}n^b,\qquad \mu,\nu,\rho=0,1,2, \quad a,b=1,2,
\ea
\ee
where $\partial_{\nu}=\frac{\partial}{\partial x^{\nu}}$ and $x^{\nu}=(\tau, r_h, \Theta)$. The unit vector $n$ is given by $n=(n^1,n^2)$, where $n^1=\frac{\phi^{r_h}}{||\phi||}$ and $n^2=\frac{\phi^{\Theta}}{||\phi||}$. It is easy to show that the topological current is conserved, i.e. $\partial_{\mu} j^{\mu}=0$. Following the $\phi$-mapping theory, it can then be proved that~\cite{AZ,BA,BB} (See Appendix for the details)
\be
\ba
\label{eq:14}
j^{\mu}=\delta^2(\phi)J^{\mu}(\frac{\phi}{x}),
\ea
\ee
where the vecor Jacobi is given by 
\be
\ba
\label{eq:777}
{\epsilon}^{ab}J^{\mu}(\frac{\phi}{x})={\epsilon}^{\mu\nu\rho}\partial_{\nu}\phi^a\partial_{\rho}\phi^b. 
\ea
\ee
When $\mu=0$, the vector Jacobi recovers the usual Jacobi as $J^0(\frac{\phi}{x})=\frac{\partial(\phi^1,\phi^2)}{\partial(x^1,x^2)}$. The equation~(\ref{eq:14}) shows that $j^{\mu}$ is zero except at $\phi=0$. After some calculations, we can find that the topological number or the total charge $W$ can be derived as~\cite{AZ,BA,BB} (See Appendix for the details)
\be
\ba
\label{eq:15}
W=\int_{\Sigma}j^0 d^2 x=\sum_{i=1}^{N}\beta_i\eta_i=\sum_{i=1}^{N}w_i,
\ea
\ee
where the positive integer $\beta_i$ is the Hopf index counting the number of the loops that $\phi$ makes when $x^u$ goes around the zero point $z_i$, $\eta_i=$sign$(J^0(\frac{\phi}{x})_{z_i})=\pm 1$ is the Brouwer degree, and $w_i$ is the winding number for the $i$-th zero point of $\phi$ in the whole parameter space $\Sigma$.

Let us make a brief summary about the topological current theory. Given a vector $\phi$, the normalized vector $n=\phi/||\phi||$ has singularities when $\phi=0$. These singularies can be treated as the topological defects, and a topological current $j^{\mu}$ can be constructed in Eq.~(\ref{eq:13}). The conserved current $j^{\mu}$ possesses a peculiar property that $j^{\mu}$ is nonzero only when $\phi=0$, which leads to the interesting inner structure of the total charge (or topological number) $W$ in Eq.~(\ref{eq:15}). Namely, the total charge is the sum of the contributions of $N$ isolated points $z_i$, which vanish the vector $\phi$. Furthermore, the contribution of each zero point $z_i$ is the winding number $w_i=\beta_i\eta_i$. Thus, the isolated points $z_i$ in topological current theory play the similar role of holes in topology.

The positive or negative winding number can reflect the different local topological natures, which is conjectured to be related to the thermodynamic stability, i.e. the positive or negative winding number corresponds to the stable and unstable black hole branches respectively~\cite{BF}. However, we note that such correspondence is strict, and we will give the proof as follows.

At first, we choose $a=1$, $b=2$ and $\mu=0$ in Eq.~(\ref{eq:777}), we can obtain
\be
\ba
\label{eq:317}
J^0(\frac{\phi}{x})&=\frac{\partial\phi^1}{\partial x^1}\frac{\partial\phi^2}{\partial x^2}-\frac{\partial\phi^1}{\partial x^2}\frac{\partial\phi^2}{\partial x^1}\\
&=\frac{\partial^2 G}{\partial r_h^2}(\frac{1+\cos^2\Theta}{\sin^3\Theta}),
\ea
\ee
where $(x^1, x^2)=(r_h, \Theta)$ and $(\phi^1, \phi^2)=(\frac{\partial G}{\partial r_h}, -\cot\Theta\csc\Theta)$ in our paper. Eq.~(\ref{eq:14}) tells us that $j^u$ is nonzero only when $\phi$ vanishes. On the one hand, $\phi^1=0$ will yield $\frac{\partial G}{\partial r_h}=0$, which represents the on-shell black hole solutions. On the other hand,  $\phi^2=0$ will yield $\Theta=\frac{\pi}{2}$, and then $(\frac{1+\cos^2\Theta}{\sin^3\Theta})=1$. Thus, Eq.~(\ref{eq:317}) can be rewritten as
\be
\ba
\label{eq:986}
J^0(\frac{\phi}{x})|_{z_i}=\frac{\partial^2 G}{\partial r_h^2}|_{z_i}.
\ea
\ee

Furthermore, $\delta^2(\phi)$ can be expanded as 
\be
\ba
\label{eq:987}
\delta^2(\phi)=\sum_{i=1}^{N}\frac{1}{|J^0(\frac{\phi}{x})|_{z_i}}\delta^2(x-z_i(t)),
\ea
\ee

Inserting Eq.~(\ref{eq:986}), Eq.~(\ref{eq:987}) and Eq.~(\ref{eq:14}) into Eq.~(\ref{eq:15}), we can obtain
\be
\ba
\label{eq:318}
W=\sum_{i=1}^Nsign(\frac{\partial^2 G}{\partial r_h^2})|_{z_i},
\ea
\ee
In Eq.~(\ref{eq:318}), we can find that the winding number $w_i$ for each on-shell black hole is equal to $sign(\frac{\partial^2 G}{\partial r_h^2})|_{z_i}$ in the corresponding extremal point (or the on-shell black hole) in the free energy landscape. Thus, there is always a correspondence between the sign of the winding number and the thermodynamic stability.

Furthermore, the topological number $W$ is defined as the sum of the winding number in the whole parameter space, which reflects the global topological nature and can be used to classify the black holes to different classes~\cite{BF}. In~\cite{BF}, the authors speculate that there are only three classes of black holes by means of the topological number. Although more and more examples show that such conjecture seems to be right~\cite{BF,BL,QC,QD,QF,QG}, a strict proof is still absent. In section~{\ref{WTTC}}, we give a proof in part, which is based on the condition $(\partial_{r_h}S)_P>0$. The proof beyond the condition $(\partial_{r_h}S)_P>0$ deserves the future considerations. If we arbitrarily choose the area $\Sigma$, the equation~(\ref{eq:15}) calculates a quantity named topological charge $\Upsilon$. When $\Sigma$ covers the whole parameter space, the topological charge $\Upsilon$ is actually the total charge $W$ (the topological number), which reflects the global topological nature. When $\Sigma$ only encloses a zero point $z_i$, the topological charge $\Upsilon$ is actually the winding number and reflects the local topological nature. If two loops $\partial\Sigma_1$ and $\partial\Sigma_2$ enclose the same zero points of $\phi$, they have the same topological charge. So, we can arbitrarily choose the contour enclosing the same zero point of $\phi$ to calculate the topological charge. A convenient choice of the contour $C$ is~\cite{BF}
\be
\ba
\label{eq:16}
r_h=a\cos\theta+r_0,\\
\Theta=b\sin\theta+\frac{\pi}{2},
\ea
\ee
where $0\leqslant \theta \leqslant \pi$, $a$, $b$ and $r_0$ can be arbitrarily chosen for the different needs of calculations. 

In the previous studies, the topological number of the vector $\phi$ can be calculated as~\cite{BD,BE,BF} 
\be
\ba
\label{eq:711}
W=\frac{1}{2\pi}\oint_{C}dA,
\ea
\ee
where $A=\arctan(\frac{\phi^2}{\phi^1})$.
However, the relationship between Eq.~(\ref{eq:711}) and Eq.~(\ref{eq:15}) has not been given. We will show that they are equivalent, i.e. we derive Eq.~(\ref{eq:711}) from Duan's topological current theory. 

At first, we choose $\mu=0$ in Eq.~(\ref{eq:13}), we can obtain
\be
\ba
\label{eq:666}
j^0&=\frac{1}{2\pi}\epsilon^{ij}\epsilon_{ab}\partial_{i}n^a\partial_{j}n^b\\
&=\partial_i(\frac{1}{2\pi}\epsilon^{ij}\epsilon_{ab}n^a\partial_jn^b)-\frac{1}{2\pi}\epsilon^{ij}\epsilon_{ab}n^a\partial_i\partial_jn^b\\
&=\partial_1(\frac{1}{2\pi}\epsilon_{ab}n^a\partial_2n^b)-\partial_2(\frac{1}{2\pi}\epsilon_{ab}n^a\partial_1n^b),  \qquad i,j,a,b=1,2,
\ea
\ee
where $\frac{1}{2\pi}\epsilon^{ij}\epsilon_{ab}n^a\partial_i\partial_jn^b$ vanishes due to $\epsilon^{ij}=-\epsilon^{ji}$. Inserting Eq.~(\ref{eq:666}) into Eq.~(\ref{eq:15}), we can obtain
\be
\ba
\label{eq:667}
W&=\int_{\Sigma}\partial_1(\frac{1}{2\pi}\epsilon_{ab}n^a\partial_2n^b)-\partial_2(\frac{1}{2\pi}\epsilon_{ab}n^a\partial_1n^b) dx^1dx^2 \\
&=\frac{1}{2\pi}\oint_C \epsilon_{ab}n^a\partial_in^b dx^i\\
&=\frac{1}{2\pi}\oint_CdA,
\ea
\ee
where we have used the divergence theorem and $A=\arctan(\frac{n^2}{n^1})=\arctan(\frac{\phi^2}{\phi^1})$. Then, a quantity which measures the angle deflection of the vector $\phi$ along the given contour is introduced by~\cite{BD,BE,BF}
\be
\ba
\label{eq:17}
\Omega(\theta)=\int_0^{\theta}\eta_{ab}n^a\partial_{\theta}n^b d\theta.
\ea
\ee
where the topology charge obeys~\cite{BD,BE}
\be
\ba
\label{eq:1000}
\Upsilon=\frac{\Omega(2\pi)}{2\pi}.
\ea
\ee

\section{When the charge is present}
\label{WCP}
\subsection{$k=0$}
At first, we discuss the simplest case when the horizon is Ricci flat. The generalized Gibbs free energy in the case $k=0$ can be written as
\be
\ba
\label{eq:18}
G=\frac{P{r_h}^{d-1}}{d-1}+\frac{Q^2}{8\pi(d-3){r_h}^{d-3}}-\frac{{r_h}^{d-2}}{4\tau}.
\ea
\ee

Then the component $\phi^{r_h}$ can be calculated as 
\be
\ba
\label{eq:19}
\phi^{r_h}=\frac{\partial G}{\partial r_h}=P{r_h}^{d-2}-\frac{Q^2}{8\pi{r_h}^{d-2}}-\frac{d-2}{4\tau}{r_h}^{d-3}.
\ea
\ee

By solving $\phi^{r_h}=0$, we can obtain
\be
\ba
\label{eq:20}
\tau=\frac{2\pi(d-2)}{8\pi P r_h-\frac{Q^2}{{r_h}^{2d-5}}}.
\ea
\ee

\begin{figure}[t]
\centering
\includegraphics[width=0.48\textwidth]{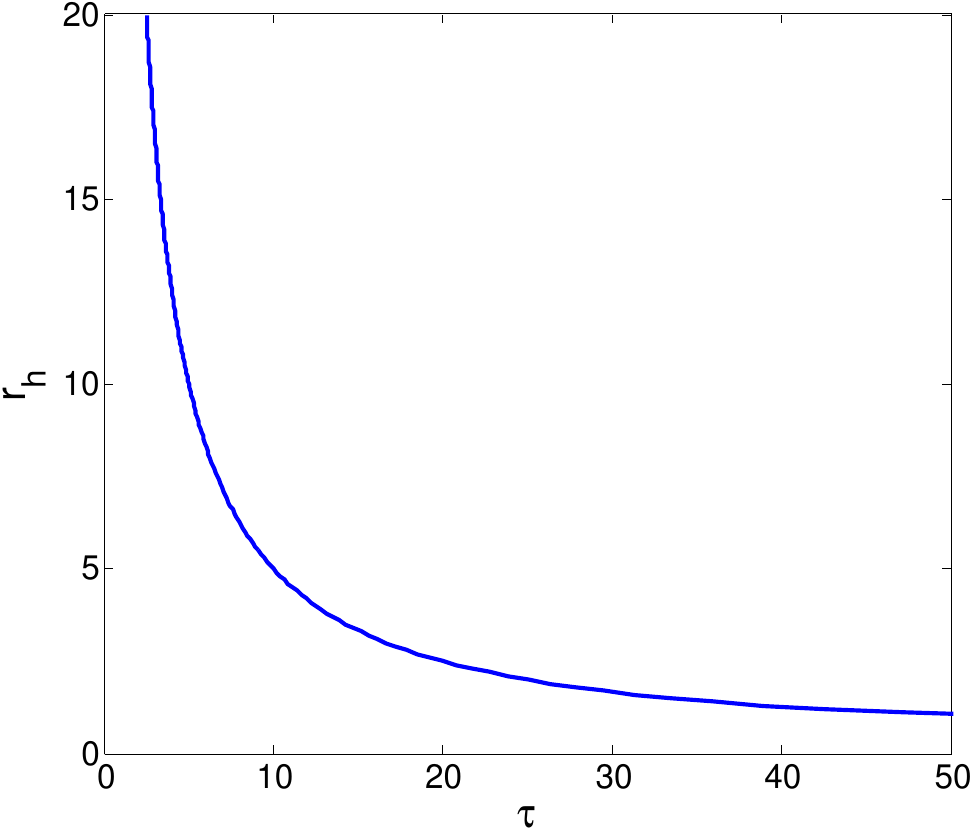}
\caption{The curve of equation~(\ref{eq:20}) in $k=0$ with $d=6$, $Q=0.2$ and $P=0.02$. Each point in the curve corresponds to an on-shell black hole.}
\label{Fig1}
\end{figure}
When we fix the ensemble temperature $T$, the Euclidean time period $\tau$ is then fixed by $\tau=\frac{1}{T}$. As discussed in the Introduction and section~\ref{TAT}, the extremal points of the Gibbs free energy landscape correspond to the on-shell black hole branches, where the ensemble temperature is equal to the Hawking temperature. So, the number of the solutions of the equation~(\ref{eq:20}) in a fixed $\tau$ is equal to the number of on-shell black holes in a fixed temperature $T$. Obviously, $\tau$ monotonically decreases with the horizon radius $r_h$, which implies that there is only one on-shell black hole in arbitrarily fixed temperature and no phase transition can occur. Furthermore, we have analyzed that the sign of the heat capacity $C_P$ is same as the sign of $(\partial_{r_h}T_H)_P$ in section~\ref{TAT}, which means $C_P$ is positive and the only one black hole is thermodynamic stable. We should note such nature is independent of the dimensions.

\begin{figure}[t]
\centering
\includegraphics[width=0.96\textwidth]{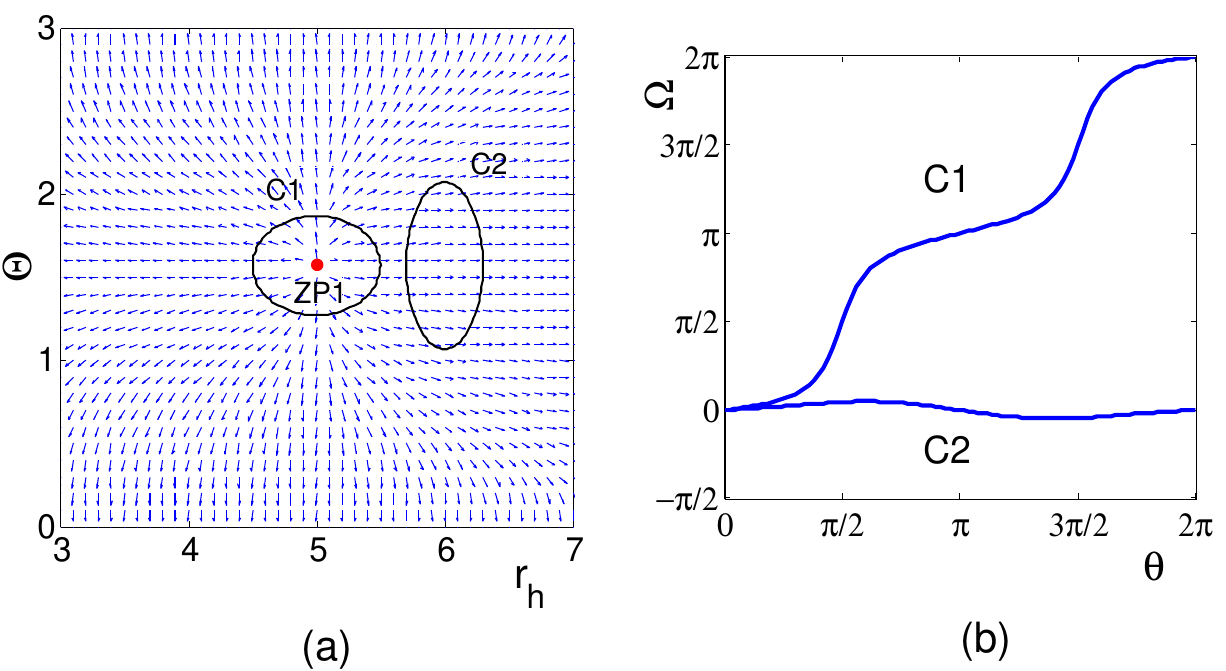}
\caption{The figures are plotted in $d=6$, $Q=0.2$, $P=0.02$ and $\tau=10$. The parameters of $C1$ and $C2$ are $a=0.5$, $b=0.3$, $r_0=5$ and $a=0.3$, $b=0.5$, $r_0=6$. $(a)$ The blue arrows show the unit vector field $n$, and the red point $ZP1$ locating at $(5,\pi/2)$ represents the zero point of $\phi$. $(b)$ The deflection $\Omega$ as a function of $\theta$ for contours $C1$ and $C2$ respectively.}
\label{Fig2}
\end{figure}

In figure~\ref{Fig1}, we have plotted the curve of equation~(\ref{eq:20}) at $d=6$, $Q=0.2$ and $P=0.02$, where the pressure satisfies the constrain~(\ref{eq:7}). From the figure, we can clearly see that there is only one on-shell black hole for an arbitrarily fixed $\tau$. Without loss of generality, we choose $\tau=10$ to study the topological properties. In figure~\ref{Fig2}.$(a)$, we have plotted the unit vector field $n$. We can observe only one zero point $ZP1$ which is located at $r_h=5$ and $\Theta=\pi/2$. For the contours $C1$ and $C2$ in figure~\ref{Fig2}.$(a)$, they are plotted by equation~(\ref{eq:16}) with $a=0.5$, $b=0.3$, $r_0=5$ and $a=0.3$, $b=0.5$, $r_0=6$. Then, the deflections of the vector field $n$ along the contours $C1$ and $C2$ can be calculated by equation~(\ref{eq:17}), which have been shown in the figure~\ref{Fig2}.$(b)$. By means of the equation~(\ref{eq:1000}), we can calculate the topological charge $\Upsilon$. The results show that the contour $C1$ enclosing the zero point $ZP1$ gives a topological charge $1$, and the contour $C2$ which does not enclose $ZP1$ gives a topological charge $0$. The positive winding number means that the on-shell black hole is thermodynamic stable, which is consistent with our analysis from the point of heat capacity $C_P$. Because there is only one on-shell black hole, the topological number $W$ is equal to the winding number $w$ as $W=w=1$. For the different dimensions, there is always only one stable on-shell black hole, which means the topological number is always equal to $1$ without dependence of the dimensions. Such results suggest that the topological number should be independent on the dimensions.

\subsection{$k=-1$}
In the case of $k=-1$, the generalized Gibbs free energy is given by
\be
\ba
\label{eq:21}
G=\frac{(d-2){r_h}^{d-3}}{16\pi}(-1+\frac{1}{{r_h}^2}+\frac{16\pi P{r_h}^2}{(d-1)(d-2)})+\frac{Q^2}{8\pi(d-3){r_h}^{d-3}}-\frac{{r_h}^{d-2}}{4\tau}(1-\frac{2(d-2)}{(d-4){r_h}^2}).
\ea
\ee

By solving $\phi^{r_h}=\frac{\partial G}{\partial r_h}=0$, we can obtain
\be
\ba
\label{eq:22}
\tau=&\frac{4\pi(d-2){r_h}^{d-3}-8\pi(d-2){r_h}^{d-5}}{-(d-2)(d-3){r_h}^{d-4}+(d-2)(d-5){r_h}^{d-6}+16\pi P{r_h}^{d-2}-2Q^2{r_h}^{2-d}}\\
=&\frac{4\pi(d-2)r_h({r_h}^2-2)}{-(d-2)(d-3){r_h}^2+(d-2)(d-5)+16\pi P{r_h}^4-\frac{2Q^2}{{r_h}^{2d-8}}},
\ea
\ee

\begin{figure}[t]
\centering
\includegraphics[width=0.48\textwidth]{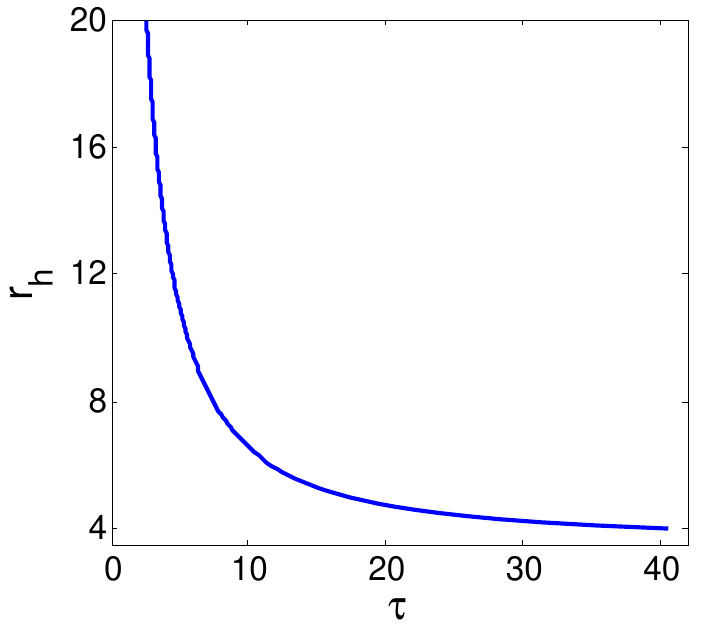}
\caption{The curve of equation~(\ref{eq:22}) in $k=-1$ with $d=6$, $P=0.02$ and $Q=0.2$. Each point in the curve corresponds to an on-shell black hole.}
\label{Fig3}
\end{figure}

\begin{figure}[t]
\centering
\includegraphics[width=0.96\textwidth]{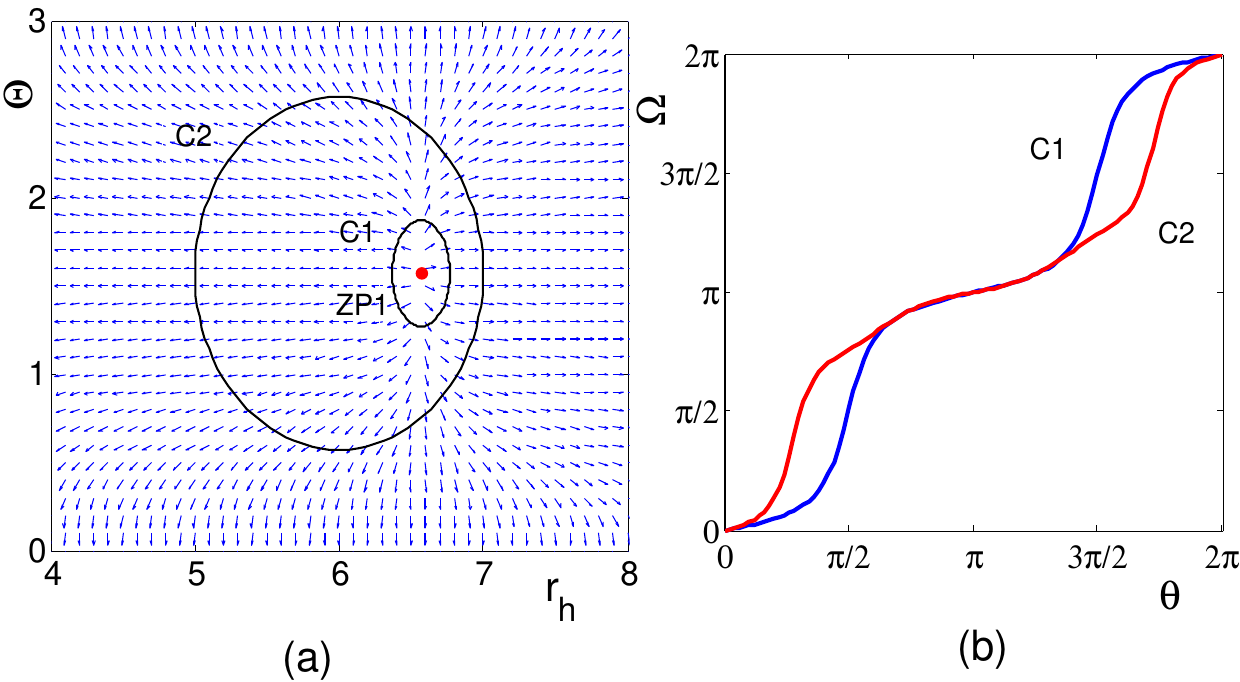}
\caption{The figures are plotted in $d=6$, $Q=0.2$, $P=0.02$ and $\tau=10$. The parameters of $C1$ and $C2$ are $a=0.2$, $b=0.3$, $r_0=6.57$ and $a=1$, $b=1$, $r_0=6$. $(a)$ The blue arrows show the unit vector field $n$, and the red point $ZP1$ locating at $(6.57, \pi/2)$ represents the zero point of $\phi$. $(b)$ The deflection $\Omega$ as a function of $\theta$ for contours $C1$ and $C2$ respectively.}
\label{Fig4}
\end{figure}

In~\cite{AP}, it was found that there is no critical point or phase transition in the case of $d=-1$. The monotonicity of the curve $\tau(r_h)$ is not easy to analyze and may depend on the parameters. Thus we plot the curve $\tau(r_h)$ under fixed parameters. In figure~\ref{Fig3}, we have plotted the curve of equation~(\ref{eq:22}) at $d=6$, $Q=0.2$ and $P=0.02$. From the figure, we can clearly see that $\tau$ monotonically decreases with the horizon radius $r_h$, which implies that there is only one on-shell black hole for an arbitrarily fixed $\tau$ and no phase transition to occur. The sign of $(\partial_{r_h}T_H)_P$ is positive, so the heat capacity is also positive and the black hole is stable. Without loss of generality, we set $\tau=10$ to study the topological properties. In figure~\ref{Fig4}.$(a)$, we have plotted the unit vector field $n$. We can observe that only one zero point $ZP1$ which is located at $r_h=6.57$ and $\Theta=\pi/2$. The contours $C1$ and $C2$ in the figure~\ref{Fig4}.$(a)$ are plotted by equation~(\ref{eq:16}) with $a=0.2$, $b=0.3$, $r_0=6.57$ and $a=1$, $b=1$, $r_0=6$ respectively. In the figure~\ref{Fig4}.$(b)$, we have plotted the deflections along the contours $C1$ and $C2$. Then, the topological charges $\Upsilon$ of contours $C1$ and $C2$ can be calculated as the same to $1$, which means that the different contours enclosing the same zero point of $\phi$ possess the same topological charge. Because there is only one zero point, the winding number $w$ is equal to the topological number $W$ as $1$.

\subsection{$k=1$}
In this section, we will discuss the case of $k=1$. In the previous studies, it was found that there is only SBH/LBH phase transition for $d=5$ and $d\geq 7$~\cite{AQ}. However, $d=6$ is an exception, where the triple point, the SBH/LBH and SBH/IBH/LBH phase transitions can be found~\cite{AQ}. Thus, we wonder whether such an exception will change the global topological nature, i.e the topological number. Because the phase behaviors of $d=5$ and $d\geq 7$ are similar, we only discuss the case of $d=5$ and divide the section into two subsection of $d=5$ and $d=6$.

In the case of $k=1$, the generalized Gibbs free energy is given by
\be
\ba
\label{eq:1100}
G=\frac{(d-2){r_h}^{d-3}}{16\pi}(1+\frac{1}{{r_h}^2}+\frac{16\pi P{r_h}^2}{(d-1)(d-2)})+\frac{Q^2}{8\pi(d-3){r_h}^{d-3}}-\frac{{r_h}^{d-2}}{4\tau}(1+\frac{2(d-2)}{(d-4){r_h}^2}).
\ea
\ee
By solving $\phi^{r_h}=\frac{\partial G}{\partial r_h}=0$, we can obtain
\be
\ba
\label{eq:1101}
\tau=\frac{4\pi (d-2) r_h ({r_h}^2+2)}{(d-2)(d-3){r_h}^2+(d-2)(d-5)+16\pi P{r_h}^4-\frac{2Q^2}{{r_h}^{2d-8}}}.
\ea
\ee

\subsubsection{d=5}
\label{subsd5}
In the case of $d=5$, the generalized Gibbs free energy is given by
\be
\ba
\label{eq:26}
G=\frac{3{r_h}^2}{16\pi}(1+\frac{1}{{r_h}^2}+\frac{4\pi P{r_h}^2}{3})+\frac{Q^2}{16\pi{r_h}^{2}}-\frac{{r_h}^{3}}{4\tau}(1+\frac{6}{{r_h}^2}),
\ea
\ee
and the on-shell black holes satisfy
\be
\ba
\label{eq:27}
\tau=\frac{6\pi{r_h}^5+12\pi{r_h}^3}{8\pi P{r_h}^6+3{r_h}^4-Q^2}.
\ea
\ee

\begin{figure}[t]
\centering
\includegraphics[width=0.96\textwidth]{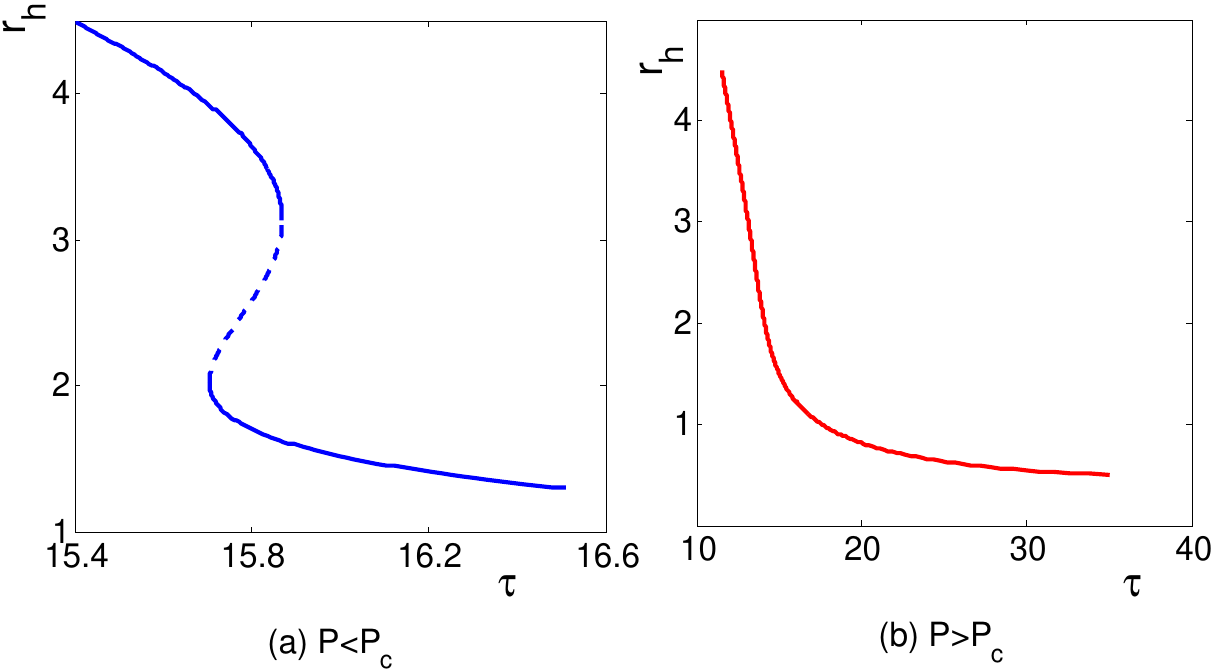}
\caption{The curves of equation~(\ref{eq:27}) in $k=1$ and $d=5$, where the stable and unstable black hole branches are plotted in solid lines and dashed line respectively. The left figure $(a)$ is plotted with $Q=0.2$ and $P=0.006<P_c$. The right figure $(b)$ is plotted with $Q=0.2$ and $P=0.01>P_c$. }
\label{Fig5}
\end{figure}

\begin{figure}[t]
\centering
\includegraphics[width=0.96\textwidth]{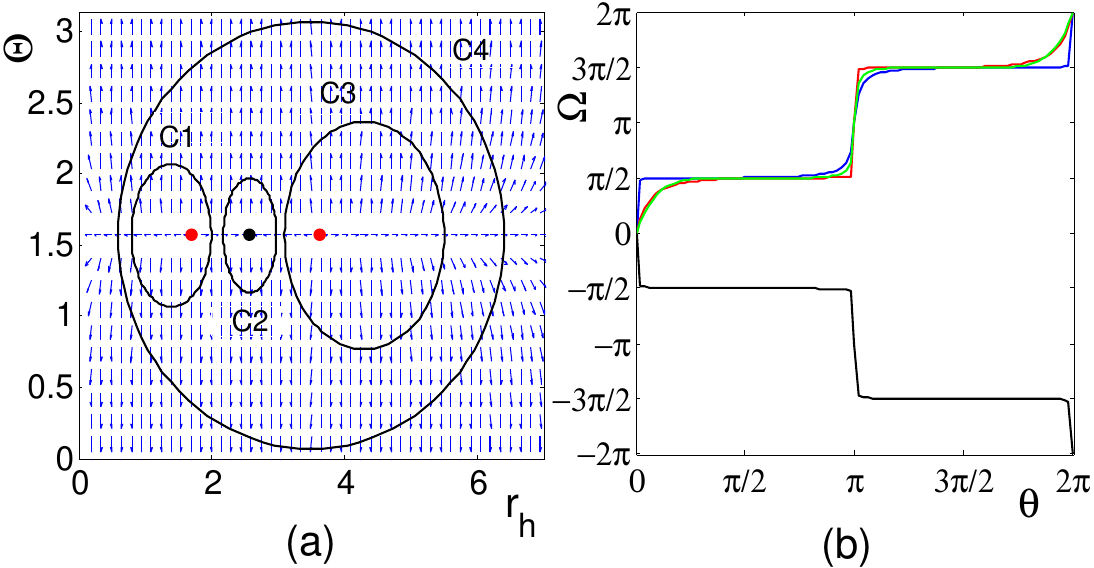}
\caption{The figures are plotted in $d=5$, $Q=0.2$, $P=0.006<P_c$ and $\tau=15.8$. The parameters of $C1$, $C2$, $C3$ and $C4$ are $a=0.6$, $0.4$, $1.2$, $2.9$, $b=0.5$, $0.4$, $0.8$, $1.5$, and $r_0=1.4$, $2.58$, $4.3$, $3.5$. $(a)$ The blue arrows show the unit vector field $n$, and the points represent the zero points of $\phi$. The points from left to right are $(1.70, \pi/2)$, $(2.58, \pi/2)$ and $(3.64,\pi/2)$, where the red points represent the stable on-shell black holes and the black point represents the unstable on-shell black hole. $(b)$ The deflection $\Omega$ as a function of $\theta$ for contours $C1$, $C2$, $C3$ and $C4$, where the blue, black, red and green curves represent $C1$, $C2$, $C3$ and $C4$ respectively.}
\label{Fig6}
\end{figure}

In~\cite{AQ}, it was found that there is a critical pressure $P_c$ depending on the charge $Q$. When $P<P_c$, the SBH/LBH phase transition can occur. However, there is no phase transition when $P>P_c$. We choose $Q=0.2$ and plot the curves of equation~(\ref{eq:27}) for $P=0.006<P_c$ and $P=0.01>P_c$ in figure~\ref{Fig5}. From the figure, we can see there are three on-shell black hole branches with two stable and one unstable black holes in certain region of $\tau$ when $P<P_c$, but there is only one stable black hole for all $\tau$ when $P>P_c$. In figure~\ref{Fig6}, we have plotted the unit vector field and the deflections along the contours $C1$, $C2$, $C3$ and $C4$ when $P<P_c$. The topological charges of $C1$, $C2$, $C3$ and $C4$ are $1$, $-1$, $1$ and $1$, where the positive winding number $1$ of $C1$ and $C3$ represents the stable black hole branches and the negative winding number $-1$ of $C2$ represents the unstable black hole branch. The topological charge of $C4$ is the topological number, which is equal to the sum of all the winding numbers as $W=1+(-1)+1=1$. An similar calculation of the topological number can also be applied to the region in $P>P_c$, which shows the topological numbers in $P<P_c$ and $P>P_c$ are the same as $1$. The results suggest that the topological number is independent on the value of positive pressure $P$, even though there are different phase behaviours in different region of $P$.

\subsubsection{d=6}
The charged Gauss-Bonnet AdS black hole in $d=6$ dimensions is very different from those in other dimensions. The generalized Gibbs free energy in $d=6$ is written as
\be
\ba
\label{eq:28}
G=\frac{{r_h}^3}{4\pi}(1+\frac{1}{{r_h}^2}+\frac{4\pi P{r_h}^2}{5})+\frac{Q^2}{24\pi{r_h}^{3}}-\frac{{r_h}^{4}}{4\tau}(1+\frac{4}{{r_h}^2}),
\ea
\ee
and the on-shell black holes satisfy
\be
\ba
\label{eq:29}
\tau=\frac{8\pi{r_h}^7+16\pi{r_h}^5}{6{r_h}^6+2{r_h}^4+8\pi P{r_h}^8-Q^2}.
\ea
\ee

\begin{figure}[t]
\centering
\includegraphics[width=0.96\textwidth]{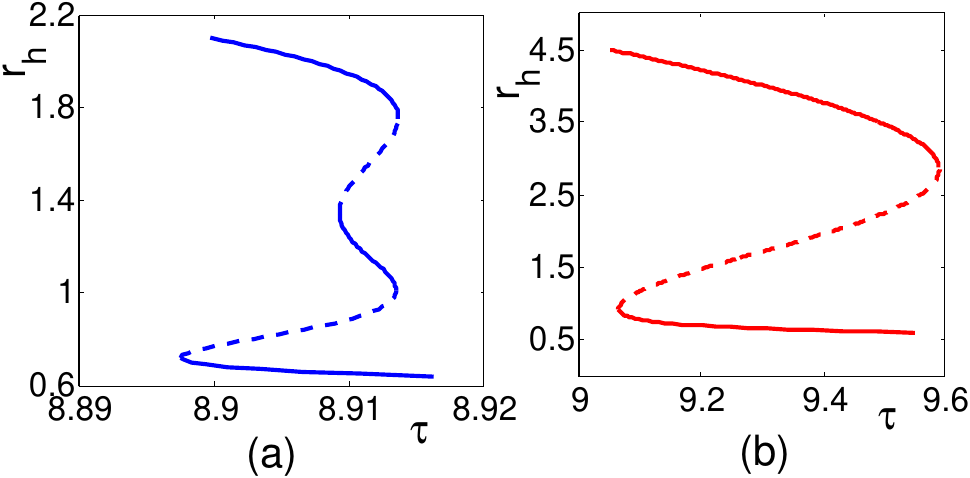}
\caption{The curves of equation~(\ref{eq:29}) in $k=1$ and $d=6$, where the stable and unstable black hole branches are plotted in solid lines and dashed line respectively. The left figure $(a)$ is plotted with $Q=0.18<Q_B$ and $P_{c1}=0.01927<P=0.01955<P_{c2}=0.01974$. The right figure $(b)$ is plotted with $Q=0.3>Q_B$ and $P=0.015<P_c=0.01956$. Beyond the regions of $r_h$ plotted in $(a)$ and $(b)$, the curves are monotonic. }
\label{Fig7}
\end{figure}

In~\cite{AQ}, it was found that there are three critical points for $Q<Q_B=0.2018$ and only one critical point for $Q>Q_B=0.2018$. For $Q<Q_B$, we denoted the pressures of the three critical points as $P_{c1}$, $P_{c2}$ and $P_{c3}$, which are dependent on the charge $Q$. When $P<P_{c1}$, there are at most three on-shell black holes. When $P_{c1}<P<P_{c2}$, there are at most five on-shell black holes. When $P_{c2}<P<P_{c3}$, there are at most three on-shell black holes. When $P>P_{c3}$, there is only one on-shell black hole. For $Q>Q_B$, there is only one critical point with critical pressure $P_c$, and the critical pressure is also dependent on the charge $Q$. There are at most three on-shell black holes in $P<P_c$ and one on-shell black hole in $P>P_c$. As suggested in section~\ref{subsd5}, we believe the topological number should be independent on the values of the positive pressure $P$. Thus, we only discuss $P_{c1}<P<P_{c2}$ for $Q<Q_B$ and $P<P_c$ for $Q>Q_B$, where there are at most five and three on-shell black holes respectively.

\begin{figure}[t]
\centering
\includegraphics[width=0.96\textwidth]{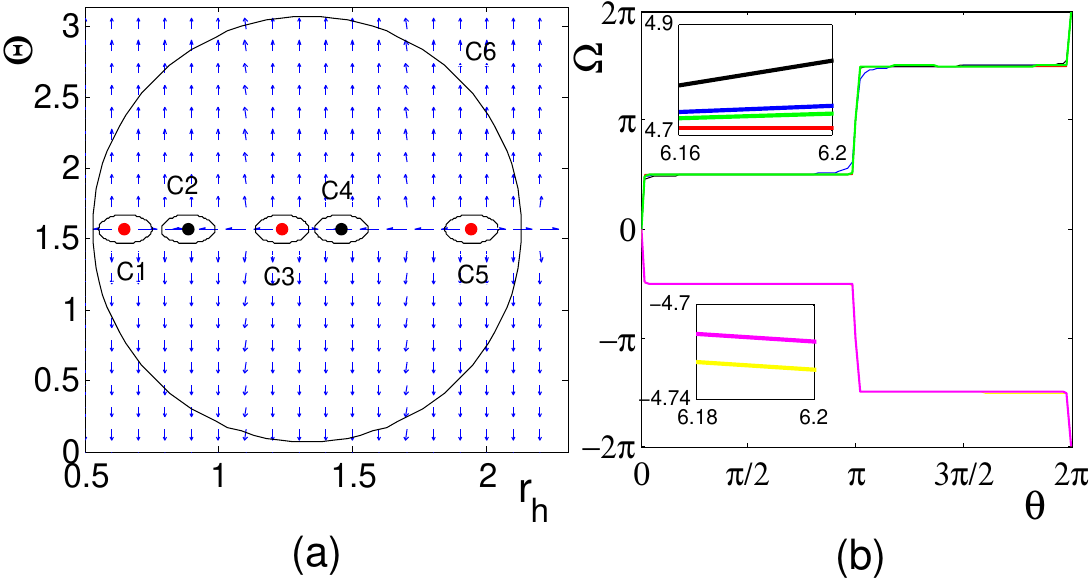}
\caption{The figures are plotted in $d=6$, $Q=0.18$, $P_{c1}=0.01927<P=0.01955<P_{c2}=0.01974$ and $\tau=8.91$. The parameters $a$ and $b$ of $C1$, $C2$, $C3$, $C4$ and $C5$ are chosen as $0.1$, $r_0$ is $0.65$, $0.89$, $1.24$, $1.46$ and $1.94$. For $C6$, $a$, $b$ and $r_0$ are $0.8$, $1.5$ and $1.33$. $(a)$ The blue arrows show the unit vector field $n$, and the points represent the zero points of $\phi$. The points from left to right are $(0.65, \pi/2)$, $(0.89, \pi/2)$, $(1.24,\pi/2)$, $(1.46,\pi/2)$ and $(1.94,\pi/2)$, where the red points represent the stable on-shell black holes and the black points represent the unstable on-shell black holes. $(b)$ The deflection $\Omega$ as a function of $\theta$ for contours $C1$, $C2$, $C3$, $C4$, $C5$ and $C6$, where the blue, yellow, red, magenta, black and green curves represent $C1$, $C2$, $C3$, $C4$, $C5$ and $C6$ respectively.}
\label{Fig8}
\end{figure}

\begin{figure}[t]
\centering
\includegraphics[width=0.96\textwidth]{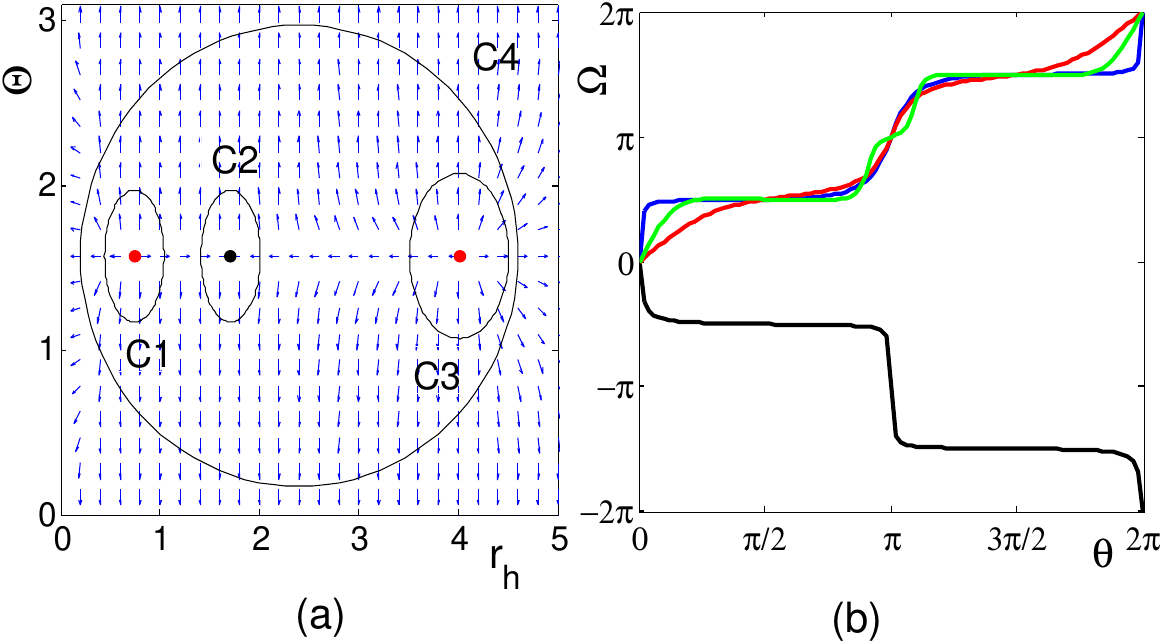}
\caption{The figures are plotted in $d=6$, $Q=0.3>Q_B$, $P=0.015<P_c=0.01956$ and $\tau=9.3$. The parameters of $C1$, $C2$, $C3$ and $C4$ are chosen as $a=0.3$, $0.3$, $0.5$, $2.2$, $b=0.4$, $0.4$, $0.5$, $1.4$ and $r_0=0.74$, $1.71$, $4.01$, $2.4$. $(a)$ The blue arrows show the unit vector field $n$, and the points represent the zero points of $\phi$. The points from left to right are $(0.74, \pi/2)$, $(1.71, \pi/2)$ and $(4.01,\pi/2)$, where the red points represent the stable on-shell black holes and the black point represents the unstable on-shell black hole. $(b)$ The deflection $\Omega$ as a function of $\theta$ for contours $C1$, $C2$, $C3$ and $C4$, where the blue, black, red and green curves represent $C1$, $C2$, $C3$ and $C4$ respectively.}
\label{Fig9}
\end{figure}

In figure~\ref{Fig7}, we have plotted the curves of equation~(\ref{eq:29}) for $Q<Q_B$, $P_{c1}<P<P_{c2}$ in $(a)$ and $Q>Q_B$, $P<P_c$ in $(b)$. From the figure, we can see that there are at most five on-shell black holes with three stable and two unstable in $(a)$ and three on-shell black holes with two stable and one unstable in $(b)$. In order to study their local and global topological topological natures, we choose $\tau=8.91$ in figure~\ref{Fig7} $(a)$ and $\tau=9.3$ in figure~\ref{Fig7} $(b)$, such that there are five  and three on-shell black holes branches respectively. In figure~\ref{Fig8} and figure~\ref{Fig9}, we have plotted the unit vector fields and the deflections along the different contours. For figure~\ref{Fig8}, it shows that the topological charges of the contours $C1$, $C2$, $C3$, $C4$ and $C5$ are $1$, $-1$, $1$, $-1$ and $1$, which is consistent with the speculation that the positive or negative winding number corresponds to the stable and unstable black hole branches. The topological charge of the contour $C6$ is the topological number, which is equal to the sum of all the winding numbers as $W=1+(-1)+1+(-1)+1=1$. For figure~\ref{Fig9}, it shows that the topological charges of the contours $C1$, $C2$ and $C3$ are $1$, $-1$ and $1$, with the positive or negative winding numbers correspond to the stable and unstable black holes branches. The topological charge of the contours $C4$ is the topological number, which also satisfies $W=1+(-1)+1+1=1$. The results suggest that the values of non-zero charge will not change the topological number, even though there are different phase behaviors in different region of the charge. Furthermore, $d=6$ is an exception in the phase transition of charged Gauss-Bonnet AdS black hole with spherical horizon, however, the topological number in $d=6$ is the same as that in $d=5$, and the topological number is still independent of the dimensions.

\section{When the charge is absent}
\label{WCA}
In~\cite{BF}, it was suggested that the different black holes possessing the same vanishing/diverging behaviours of the curves $\tau({r_h})$ at the small $r_h$ and large $r_h$ limits have the same topological number, and can be divided to the same class. The equivalence between the asymptotic behaviour and the topological number is a conjecture, and the proof needs the future considerations. At first, we examine such conjecture in the cases discussed in the previous sections, i.e. when the charge is present. Then, we will discuss the case when the charge is absent.

When the charge is present, we should note the lower limits of $r_h$ in equations~(\ref{eq:20}),~(\ref{eq:22}) and ~(\ref{eq:1101}) are $r_{ex}$ rather than zero, where $r_{ex}$ represents the horizon radius of the extremal black hole with zero temperature. Although the lower bound of $r_h$ of the vector field $n$ is always zero, $r_h$ in the equation $\tau$ represents the radius of a physical on-shell black hole with non-negative temperature. When the temperature is zero, it is exactly the extremal black hole. It is easy to find that $\tau$ possesses the same asymptotic behaviours in $k=0$, $k=-1$ and $k=1$ as
\be
\ba
\label{eq:30}
&\tau\to\infty \qquad r_h\to r_{ex},\\
&\tau\to 0 \qquad r_h\to\infty.
\ea
\ee
It coincides with our previous calculations that the black holes in $k=0$, $k=-1$ and $k=1$ possess the same topological number $1$, so those black holes can be divided into the same class as RNAdS black hole~\cite{BF}. We should note that such asymptotic behaviours are independent on the dimensions, positive pressure and non-zero charge, just the same as we have discussed about the topological number in the previous section.

If the charge is absent, the equation of $\tau$~(\ref{eq:20}) in $k=0$ is rewritten as 
\be
\ba
\label{eq:31}
\tau=\frac{2\pi (d-2)}{8\pi P r_h}.
\ea
\ee
$\tau$ is still monotonically diminishing with $r_h$, so the heat capacity $C_P$ is still positive and there is only one on-shell stable black hole. Therefore, the winding number is equal to the topological number as $1$. The lower bound of $r_h$ in the equation~(\ref{eq:31}) is zero, and the asymptotic behaviours of $\tau$ in small and large $r_h$ limit are shown as
\be
\ba
\label{eq:32}
&\tau\to\infty \qquad r_h\to 0,\\
&\tau\to 0 \qquad r_h\to\infty,
\ea
\ee
which are same as the cases when the charge is present. 

\begin{figure}[t]
\centering
\includegraphics[width=0.48\textwidth]{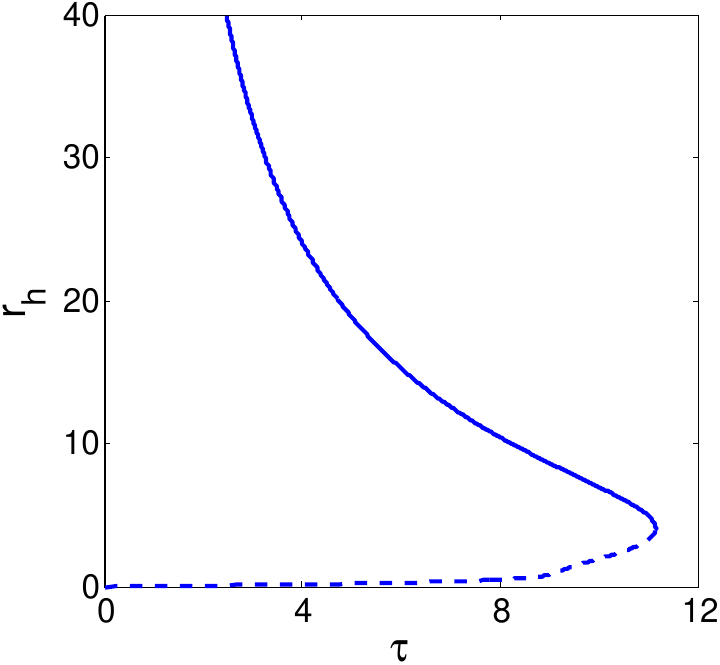}
\caption{The curves of equation~(\ref{eq:34}) in $k=1$, $d=6$, $P=0.01$, where the stable and unstable black hole branches are plotted in solid lines and dashed line respectively.}
\label{Fig10}
\end{figure}

For the case of zero charge in $k=-1$, the equation of $\tau$~(\ref{eq:22}) can be rewritten as
\be
\ba
\label{eq:33}
\tau=\frac{4\pi (d-2)r_h({r_h}^2-2)}{16\pi P{r_h}^4-(d-2)(d-3){r_h}^2+{(d-2)(d-5)}},
\ea
\ee
where $\tau=\infty$ (or $T_H=0$) can be satisfied if $P<\frac{(d-2)(d-3)^2}{64\pi (d-5)}$. It is easy to find that$\frac{(d-3)^2}{d-5}>d-1$, so there is always a lower bound $r_{ex}$ corresponding to the extremal black hole for the pressure satisfying constraint~(\ref{eq:7}). Then the asymptotic behaviours of $\tau$ in $r_h\to r_{ex}$ and $r_h\to\infty$ limits can be calculated and were found to be the same as those in the previous cases.

\begin{figure}[t]
\centering
\includegraphics[width=0.96\textwidth]{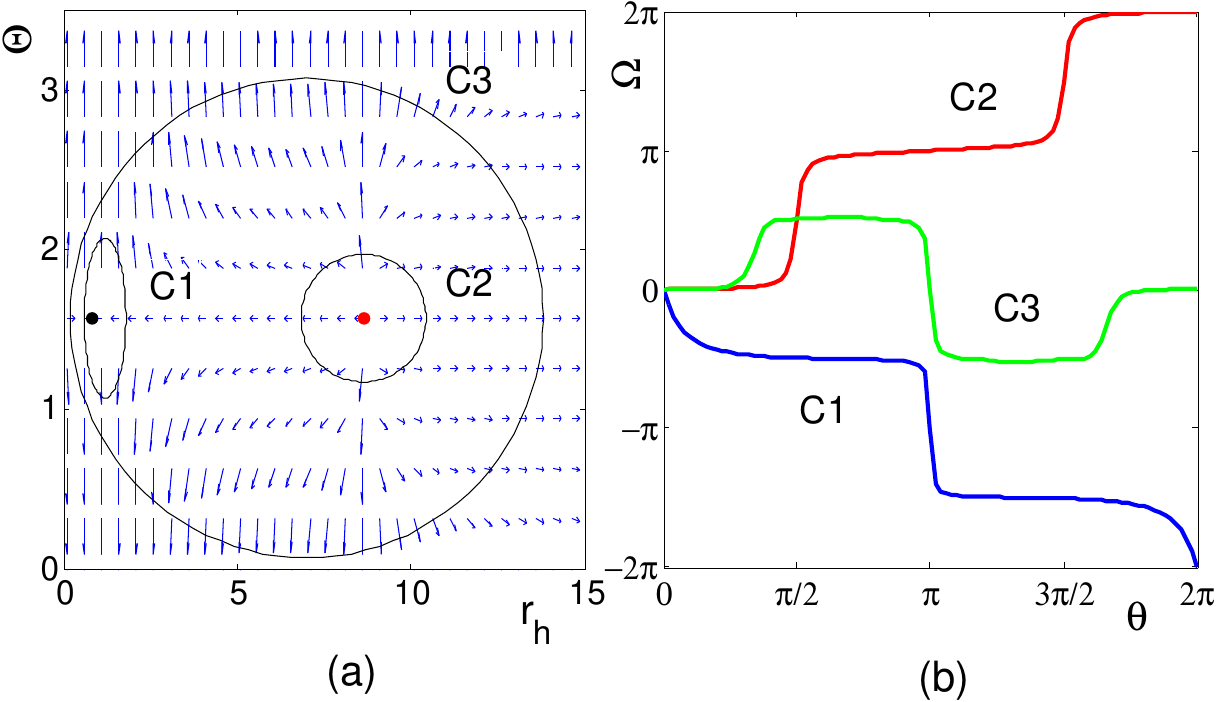}
\caption{The figures are plotted in $d=6$, $Q=0$, $P=0.01$ and $\tau=9$. The parameters of $C1$, $C2$ and $C3$ and $C4$ are chosen as $a=0.6$, $1.8$, $6.8$, $b=0.5$, $0.4$, $1.5$, and $r_0=1.2$, $8.6$, $7.0$. $(a)$ The blue arrows show the unit vector field $n$, and the points represent the zero points of $\phi$. The points from left to right are $(0.85, \pi/2)$ and $(8.63,\pi/2)$, where the red point represents the stable on-shell black hole and the black point represents the unstable on-shell black hole. $(b)$ The deflection $\Omega$ as a function of $\theta$ for contours $C1$, $C2$ and $C3$ , where the blue, red and green curves represent $C1$, $C2$ and $C3$ respectively.}
\label{Fig11}
\end{figure}

For the case of zero charge in $k=1$, the equation of $\tau$~(\ref{eq:1101}) is rewritten as
\be
\ba
\label{eq:34}
\tau=\frac{4\pi (d-2) r_h ({r_h}^2+2)}{(d-2)(d-3){r_h}^2+(d-2)(d-5)+16\pi P{r_h}^4},
\ea
\ee
where the lower bound of $r_h$ is always zero. Obviously, the asymptotic behaviours of $\tau$ in $d\geq 6$ are different from $d=5$. When $d=5$, $\tau\to\infty$ for $r_h\to 0$ and $\tau\to 0$ for $r_h\to\infty$, which is the same as the previous cases. However, for $d\geq 6$, we can obtain
\be
\ba
\label{eq:35}
&\tau\to 0 \qquad r_h\to 0,\\
&\tau\to 0 \qquad r_h\to\infty,
\ea
\ee
such asymptotic behaviour does not occur in the previous cases and the studies in~\cite{BF}. In figure~\ref{Fig10}, we have chosen $d=6$ and $P=0.01$ to plot the curve of $\tau(r_h)$ in equation~(\ref{eq:34}). From the figure, we can see that there are two on-shell black holes, where the one with larger radius is stable and the other one is unstable. Furthermore, we can see clearly that $\tau\to 0$ when $r_h\to 0$ and $\infty$. In figure~\ref{Fig11}, we have plotted the unit vector fields and the deflections along the different contours. The winding number of $C1$ (or $C2$) enclosing the unstable (or stable) black hole branch is $-1$ (or $1$), and the topological number satisfies $W=1+(-1)=0$. Such a topological number is equal to that of the RN black hole, but the asymptotic behaviours of $\tau$ in RN black hole behave as $\tau\to\infty$ for $r_h\to r_{ex}$ and $\tau\to \infty$ for $r_h\to\infty$~\cite{BF}. Thus, we suggest that the black holes with such two different asymptotic behaviours belong to the same topological class. There are totally three types of topological classes, which correspond to the four kinds of asymptotic behaviours as: $(1)$ $r_h\to 0$ or $\infty$, both $\tau\to 0$ or both $\tau\to\infty$, $(2)$ $\tau\to 0$ for $r_h\to 0$, $\tau\to\infty$ for $r_h\to\infty$, $(3)$ $\tau\to\infty$ for $r_h\to 0$, $\tau\to 0$ for $r_h\to\infty$.

In conclusion, if the horizon is Ricci flat or hyperbolic, all the results are the same with the cases when the charge is present. However, in the case of spherical horizon, the topological number in $d\geq 6$ dimensions is different from that in $d=5$ dimensions. When $d=5$ dimensions, the topological number is the same with the previous cases as $1$. When $d\geq 6$ dimensions, the topological number is $0$, which is equal to the topological number of the RN black holes. Actually, the differences in the phase behaviours between $d=5$ dimensions and $d\geq 6$ dimensions have been found in the previous previous studies~\cite{AP,BK}. In the case of $d=5$, the small/large black hole phase transition can always occur. While in the case of $d\geq 6$, there may be no phase transition. Although the phase behaviours can not be a criterion to distinguish the different topological numbers, while our studies show that the differences in phase behaviours between the black holes in $d=5$ and $d\geq 6$ dimensions may come from the different topologies, which are different from the other types of different phase behaviours, such as $d=6$ and $d\neq 6$ in the charged cases. The values of the non-zero charge will not change the topological number, while having zero or non-zero charge will affect the topological number. When we turn off the charge of the RN black holes, the topological number will change from $0$ to $-1$~\cite{BF}. However, our studies show the topological number will not always change when we turn off the charge, it may not necessarily change in some regions of the parameters.

In the case of the spherical horizon and zero charge, we find that the topological number shows certain dimension dependence, i.e. the differences between $d=5$ and $d\geq 6$, which is different from the previous studies. So, a specific calculation of the topological number in different black holes is required. However, a simpler method can be applied. We have examined the methods proposed in~\cite{BF}, where the asymptotic behaviours of the curve $\tau(r_h)$ in small and large radii limits can be used as a criterion to distinguish the different topological numbers, and find that it is always accurate. Furthermore, we find a new asymptotic behaviour as $\tau(r_h\to 0)=0$ and $\tau(r_h\to\infty)=0$, which does not appear in the previous studies of the black hole systems. In the next section, under the condition $(\partial_{r_h}S)_P>0$, we will give an intuitive proof of why there are only three topological classes and why two kinds of the asymptotic behaviours belong to the same topological class.


\section{Why there are only three topological classes of black holes?}
\label{WTTC}
If the condition $(\partial_{r_h}S)_P>0$ is satisfied, namely $C_P$ has the same sign of $(\partial_{r_h}T_H)_P$. Then, we can give an intuitional proof of why there are only three classes of black holes. 

\begin{figure}[t]
\centering
\includegraphics[width=0.96\textwidth]{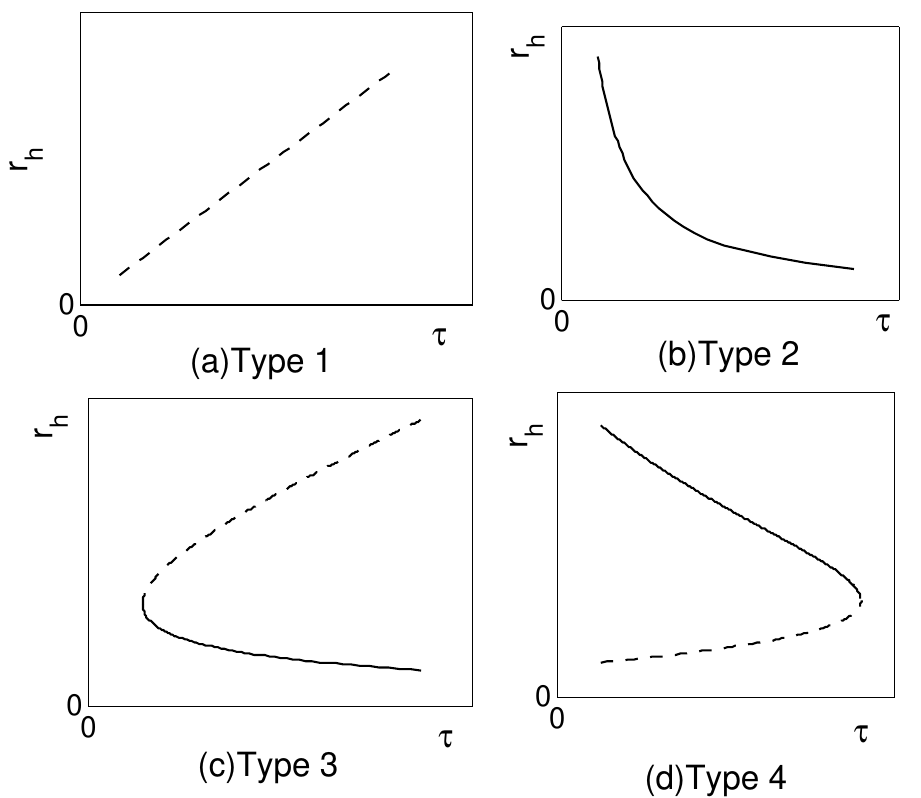}
\caption{Four types of the asymptotic behaviors, where the solid lines represent the stable on-shell black hole branches and dashes for unstable black hole branches.}
\label{Fig12}
\end{figure}

There are four types of combinations between $\tau(r_h\to r_{min})=0/\infty$ and $\tau(r_h\to\infty)=0/\infty$, where we denote the minimum value of $r_h$ as $r_{min}$, it can be $0$ or $r_{ex}$. In figure~\ref{Fig12}, we have plotted the simplest types of the curves with these four asymptotic behaviours, where the solid lines represent the stable black hole branches and the dashed lines represent the unstable black hole branches. We denote the four curves with the different asymptotic behaviours as: Type$1$. $\tau(r_h\to r_{min})=0$ and $\tau(r_h\to\infty)=\infty$, Type$2$. $\tau(r_h\to r_{min})=\infty$ and $\tau(r_h\to\infty)=0$, Type$3$. $\tau(r_h\to r_{min})=\infty$ and $\tau(r_h\to\infty)=\infty$, Type$4$. $\tau(r_h\to r_{min})=0$ and $\tau(r_h\to\infty)=0$. Because the winding numbers of the stable or unstable black holes are $1$ and $-1$ respectively, the topological numbers as the sum of the winding numbers can be calculated. For Type $1$, $W=-1$. For Type $2$, $W=1$. For Type $3$, $W=1+(-1)=0$. For Type $4$, $W=(-1)+1=0$. For different black holes, $\tau(r_h)$ is always a continuous function of $r_h$, so every $\tau(r_h)$ curve belongs to four types, 
\be
\ba
Type1+n\times Type3,\\
Type2+n\times Type4,\\
Type3+n\times Type3,\\
Type4+n\times Type4,\\
\ea
\ee
where $n$ represents the non-negative integer. Because the topological numbers of Type $3$ and Type $4$ are zero, the topological numbers of all $\tau(r_h)$ curves (or all on-shell black holes) can only be $1$, $-1$ and $0$. We note the curves of Type $3$ and Type $4$ have different asymptotic behaviors but possess the same topological number $0$, so we suggest that the black holes with the asymptotic behaviors Type $3$ $(\tau(r_h\to r_{min})=\infty,\tau(r_h\to\infty)=\infty)$ and Type $4$ $(\tau(r_h\to r_{min})=0,\tau(r_h\to\infty)=0)$ are topologically equivalent.


\section{Conclusion}
\label{CON}
In this paper, we have studied the local and global topological natures of the Gauss-Bonnet black hole in AdS space. The local topological features are reflected by the winding number, where the positive or negative winding number always corresponds to the stable and unstable black hole branches, respectively. We find that such correspondence is strict, and we have given the proof. The global topological features are reflected by the topological number, which is defined as the sum of the winding number, and can be used to classify the different black holes into different topological classes. We have also given a proof of this conjecture in part based on the condition $(\partial_{r_h} S)_P>0$.

In Table~\ref{tab1}, we have combined our results with those in~\cite{BF}, and five interesting properties can be concluded. Firstly, The higher derivative terms of curvature in the form of Gauss-Bonnet gravity will not change the topological number. This conclusion is conjectured from the fact that the RN-AdS black hole and the charged GB-AdS black hole possess the same topological number. Secondly, the topological number is conjectured to be independent on the size of $P$ and $Q$, but be dependent on whether $P$ and $Q$ are zero. Thirdly, the topological number of the charged GB-AdS black hole is found to be independent of the geometry of the black hole horizon. Fourthly, the topological number may have parameter dependence in some special cases. For $Q=0$ GB-AdS black hole with spherical horizon, the topological numbers are different between $d=5$ and $d\geq 6$. Such dimensional dependence is against for the conjecture in~\cite{BF} that the topological number is independent of the black hole parameters. Fifthly, a new asymptotic behaviour of $\tau(r_h)$ is found as $\tau\to 0$ for $r_h\to r_{min}$ and as $\tau\to 0$ for $r_h\to\infty$, which is conjectured to be topological equivalent with the asymptotic behaviour as $\tau\to\infty$ for $r_h\to r_{min}$ and as $\tau\to\infty$ for $r_h\to\infty$. Furthermore, such equivalence has been proved based on the condition $(\partial_{r_h} S)_P>0$ in our paper. The proof beyond the condition $(\partial_{r_h} S)_P>0$ deserves the future considerations.

\begin{table}[h]
\setlength{\tabcolsep}{3mm}{
\begin{center}
\begin{tabular}{c|c|c}
  \hline\hline
Black holes & Topological number & Asymptotic behaviour \\
\hline
$d=4$ Sch BH~\cite{BF}&$W=-1$&\makecell{$\tau\to 0$ for $r_h\to r_{min}$,\\ $\tau\to\infty$ for $r_h\to\infty$}  \\
\hline
\makecell{RN-AdS BH~\cite{BF} \\  Charged GB-AdS BH \\ $Q=0$,$k=0$ or $-1$ GB-AdS BH \\ $Q=0$, $k=1$, $d=5$ GB-AdS BH}

& $W=1$ & \makecell{$\tau\to\infty$ for $r_h\to r_{min}$,\\ $\tau\to 0$ for $r_h\to\infty$}  \\
\hline
{\begin{tabular}{c}
 $Q=0$,$k=1$,$d\geq 6$ GB-AdS BH  \\  \hline
  $d=4$ RN BH~\cite{BF} 
    \end{tabular}}
  &$W=0$ &
  {\begin{tabular}{c}
  $\tau\to 0$ for $r_h\to r_{min}$,\\ $\tau\to 0$ for $r_h\to\infty$  \\  \hline
 $\tau\to\infty$ for $r_h\to r_{min}$,\\ $\tau\to\infty$ for $r_h\to\infty$ 
    \end{tabular}}
  \\
  \hline\hline
\end{tabular}
\caption{The topological number $W$ and the asymptotic behaviours of $\tau(r_h)$ for the different black holes.}\label{tab1}
\end{center}}
\end{table}



\section*{Acknowledgements}
C. H. Liu thanks Ran Li, Xuanhua Wang, Hong Wang and He Wang for the helpful discussions. C. H. Liu thanks the support in part by the National Natural Science
Foundation of China Grant No. 21721003 and No. 12234019.

\section*{Appendix: The topological construction} 
In the Appendix, we will introduce the topological current theory reported in~\cite{AZ,BA,BB,BC,BF}.

Beginning with a vector field $\phi=(\phi^r,\phi^{\theta})$, the normalized vectors are defined as 
\be
\ba
\label{eq:300}
n^a=\frac{\phi^a}{||\phi||}, \qquad a=1,2,
\ea
\ee
where $\phi^1=\phi^r$, $\phi^2=\phi^{\theta}$ and $||\phi||=\sqrt{(\phi^1)^2+(\phi^2)^2}$. Then, one can introduce a superpotential as
\be
\ba
\label{eq:301}
V^{\mu\nu}=\frac{1}{2\pi}\epsilon^{\mu\nu\rho}\epsilon_{ab}n^a\partial_{\rho}n^b,\qquad \mu,\nu,\rho=0,1,2,
\ea
\ee
where $\partial_{\rho}=\frac{\partial}{\partial x^{\rho}}$ and $x^{\rho}=(t,r,\theta)$. By using the superpotential $V^{\mu\nu}$, a topological current can be defined as 
\be
\ba
\label{eq:302}
j^{\mu}=\partial_{\nu}V^{\mu\nu}=\frac{1}{2\pi}\epsilon^{\mu\nu\rho}\epsilon_{ab}\partial_{\nu}n^a\partial_{\rho}n^b,
\ea
\ee
where the term $\frac{1}{2\pi}\epsilon^{\mu\nu\rho}\epsilon_{ab}n^a\partial_{\nu}\partial_{\rho}n^b$ in $\partial_{\nu}V^{\mu\nu}$ vanishes due to $\epsilon^{\mu\nu\rho}=-\epsilon^{\mu\rho\nu}$. Note $V^{\mu\nu}=-V^{\nu\mu}$, one can find that the topological current is conserved, i.e. $\partial_{\mu}j^{\mu}=0$. Thus, the total charge (also called topological number) in the whole parameter space $\Sigma$ is obtained as
\be
\ba
\label{eq:303}
W=\int_{\Sigma}j^0d^2x.
\ea
\ee

The topological current in Eq.~(\ref{eq:302}) can be rewritten as 
\be
\ba
\label{eq:304}
j^{\mu}&=\frac{1}{2\pi}\epsilon^{\mu\nu\rho}\epsilon_{ab}\partial_{\nu}(\frac{\phi^a}{||\phi||})\partial_{\rho}(\frac{\phi^b}{||\phi||})\\
&=\frac{1}{2\pi}\epsilon^{\mu\nu\rho}\epsilon_{ab}\frac{\partial}{\partial\phi^c}(\frac{\phi^a}{||\phi||^2})\partial_{\nu}\phi^c\partial_{\rho}\phi^b\\
&=\frac{1}{2\pi}\epsilon^{\mu\nu\rho}\epsilon_{ab}\frac{\partial}{\partial\phi^c}(\frac{\partial}{\partial\phi^a}\ln{||\phi||})\partial_{\nu}\phi^c\partial_{\rho}\phi^b,
\ea
\ee
where we have used $\partial_{\nu}(\frac{\phi^a}{||\phi||})=\frac{\partial}{\partial\phi^c}(\frac{\phi^a}{||\phi||})\partial_{\nu}\phi^c$ and $\frac{\partial\ln{||\phi||}}{\partial\phi^a}=\frac{\phi^a}{||\phi||^2}$.

The vector Jacobian is defined as:
\be
\ba
\label{eq:305}
\epsilon^{ab}J^{\mu}(\frac{\phi}{x})=\epsilon^{\mu\nu\rho}\partial_{\nu}\phi^a\partial_{\rho}\phi^b, \qquad \mu,\nu,\rho=0,1,2,
\ea
\ee
where 
\be
\ba
\label{eq:306}
J^0(\frac{\phi}{x})=\partial_1\phi^1\partial_2\phi^2-\partial_2\phi^1\partial_1\phi^2=\frac{\partial(\phi^1,\phi^2)}{\partial(x^1,x^2)}
\ea
\ee
is the usual Jacobian denoted as $J(\frac{\phi}{x})$.

Inserting (\ref{eq:305}) into (\ref{eq:304}), we have
\be
\ba
\label{eq:307}
j^{\mu}=\frac{1}{2\pi}(\Delta_{\phi_a}\ln||\phi||)J^{\mu}(\frac{\phi}{x}),
\ea
\ee
where $\Delta_{\phi_a}=\frac{\partial^2}{\partial\phi^a\partial\phi^a}$. Based on the two-dimensional Laplacian Green's function in $\phi$-mapping space, we have $\Delta_{\phi_a}\ln||\phi||=2\pi\delta^2(\phi)$ and Eq.~(\ref{eq:307}) can be rewritten as 
\be
\ba
\label{eq:308}
j^{\mu}=\delta^2(\phi)J^{\mu}(\frac{\phi}{x}).
\ea
\ee
Thus, the topological current $j^{\mu}$ is nonzero only when $\phi=0$, i.e.,
\be
\ba
\label{eq:309}
\phi^a(x^0,x^1,x^2)=0, \qquad a=1,2.
\ea
\ee

Considering that there are $N$ zero points of $\phi$ and the usual Jacobian $J(\frac{\phi}{x})$ is nonzero, the solutions of Eq.~(\ref{eq:309}) can be rewritten as
\be
\ba
\label{eq:310}
&x^0=t,\\
&x^i=z^i_n(t), \qquad n=1,2,\dots, N,\quad i=1,2.
\ea
\ee

In terms of $j^{\mu}$ in Eq.~(\ref{eq:308}), we can simplify it and the characteristic property of the topological current can be revealed. On the one hand, $\delta^2(\phi)$ can be expanded as
\be
\ba
\label{eq:311}
\delta^2(\phi)=\sum_{n=1}^{N}\frac{1}{|J(\frac{\phi}{x})|_{z_n}}\beta_n\delta^2(x-z_n(t)),
\ea
\ee
where $\beta_n$ is the Hopf index which measures the numbers of the loops that $\phi^a$ makes when $x$ goes around the zero point $z_n$. On the other hand, an equation is satisfied as
\be
\ba
\label{eq:312}
\frac{J^{\mu}(\frac{\phi}{x})|_{z_n}}{J^0(\frac{\phi}{x})|_{z_n}}=\frac{dx^{\mu}}{dx^0}|_{z_n},
\ea
\ee
which can be proved as follows. At first, we rewrite the vector Jacobian in Eq.~(\ref{eq:305}) as
\be
\ba
\label{eq:313}
J^{\mu}(\frac{\phi}{x})dx^0=\frac{1}{2}\epsilon_{ab}\epsilon^{\mu\nu\rho}\partial_{\nu}\phi^a\partial_{\rho}\phi^b dx^0.
\ea
\ee
When $\mu=0$, it is obvious that Eq.~(\ref{eq:312}) is satisfied. When $\mu=1$, we consider the case that Eq.~(\ref{eq:313}) locates at $x=z_n$ and yields
\be
\ba
\label{eq:314}
J^1(\frac{\phi}{x})|_{z_n}dx^0|_{z_n}&=\frac{1}{2}\epsilon_{ab}\epsilon^{1\nu\rho}(\partial_{\nu}\phi^a)|_{z_n}(\partial_{\rho}\phi^b)|_{z_n} dx^0|_{z_n}\\
&=\frac{1}{2}\epsilon_{ab}\{-\epsilon^{012}(\partial_0\phi^adx^0)|_{z_n}(\partial_2\phi^b)|_{z_n}-\epsilon^{021}(\partial_2\phi^a)|_{z_n}(\partial_0\phi^bdx^0)|_{z_n}\} \\
&=\frac{1}{2}\epsilon_{ab}\{\epsilon^{012}(\partial_i\phi^adx^i)|_{z_n}(\partial_2\phi^b)|_{z_n}+\epsilon^{021}(\partial_2\phi^a)|_{z_n}(\partial_i\phi^bdx^i)|_{z_n}\}\\
&=\frac{1}{2}\epsilon_{ab}(\epsilon^{012}\partial_1\phi^a\partial_2\phi^b+\epsilon^{021}\partial_2\phi^a\partial_1\phi^b)|_{z_n}dx^1|_{z_n}\\
&=J^0(\frac{\phi}{x})|_{z_n}dx^1|_{z_n},
\ea
\ee
where we have used $(\partial_{\mu}\phi^adx^{\mu})|_{z_n}=0$, i.e. $(\partial_0\phi^adx^0)|_{z_n}=-(\partial_i\phi^adx^i)|_{z_n}, i=1,2$. This results from the use of the implicit function theorem in Eq.~(\ref{eq:309})~\cite{BF,BZ}. The same procedure can be applied for $\mu=2$, and Eq.~(\ref{eq:312}) can be proved. Notice $j^{\mu}$ is nonzero only when $x=z_n(t)$, we can insert Eq.~(\ref{eq:311}) and Eq.~(\ref{eq:312}) into Eq.~(\ref{eq:308}) and yield
\be
\ba
\label{eq:315}
j^{\mu}=\sum_{n=1}^{N}\eta_n\beta_n\delta^2(x-z_n(t))\frac{dx^{\mu}}{dx^0}|_{z_n},
\ea
\ee
where $\eta_n=\frac{J^0(\frac{\phi}{x})|_{z_n}}{|J^0(\frac{\phi}{x})|_{z_n}}$ is the Brouwer degree at zero point $z_n$. Thus, the topological number $W$ in Eq.~(\ref{eq:303}) can be rewritten as
\be
\ba
\label{eq:316}
W=\sum_{n=1}^{N}\eta_n\beta_n,
\ea
\ee
Equation~(\ref{eq:316}) reflects the inner structure of the total charge $W$. Namely, the total charge is the sum of the contributions of $N$ isolated zero points of $\phi$. For each zero point, its contribution to the charge is the winding number $w_n=\eta_n\beta_n$.


\end{document}